\documentclass{TestStyle}

\pdfoutput=1

\usepackage{graphicx}
\usepackage{bm}
\usepackage{amsmath,amssymb}
\usepackage{xspace}

\usepackage{textcomp}
\usepackage[T1]{fontenc} 
\renewcommand{\vec}[1]{{\bf #1}} 
\newcommand{\vecr}{\ensuremath{\vec{r}}\xspace}
\newcommand{\vecq}{\vec{q}}
\newcommand{\vecp}{\vec{p}}

\newcommand{\bb}[1]{\left(#1\right)}
\newcommand{\absv}[1]{\left|#1\right|}
\newcommand{\absvsq}[1]{\absv{#1}^2}

\newcommand{\ef}[1]{\ensuremath{\operatorname{e}^{#1}}\xspace}

\newcommand{\ket}[1]{\ensuremath{\left|#1\right\rangle}\xspace}
\newcommand{\braketop}[3]{\left\langle#1\middle|#2\middle|#3\right\rangle}

\newcommand{\integralb}[3]{\int\limits_{#1}^{#2} \! \mathrm{d} #3\,}
\newcommand{\integral}[1]{\int \! \mathrm{d} #1\,}
\newcommand{\intvol}{\integral{^3r}}

\newcommand{\aop}{\ensuremath{\hat{a}^{\phantom\dagger}}\xspace}
\newcommand{\aopd}{\ensuremath{\hat{a}^\dagger}\xspace}
\newcommand{\bop}{\ensuremath{\hat{b}^{\phantom\dagger}}\xspace}
\newcommand{\bopd}{\ensuremath{\hat{b}^\dagger}\xspace}
\newcommand{\bogu}{\ensuremath{u^{\phantom\dagger}}\xspace}
\newcommand{\bogv}{\ensuremath{v^{\phantom\dagger}}\xspace}
\newcommand{\rhoq}{\ensuremath{\rho^{\phantom\dagger}_{\vecq}}\xspace}

\title{Coupling a single electron to a Bose-Einstein condensate}
\author{Jonathan B. Balewski$^{1}$, Alexander T. Krupp$^{1}$, Anita Gaj$^{1}$, David Peter$^{2}$, Hans Peter B\"{u}chler$^{2}$, Robert L\"{o}w$^{1}$, Sebastian Hofferberth$^{1}$ \& Tilman Pfau$^{1}$}

\begin{document}

\maketitle
~\\
\vspace{-1.005cm} 
\begin{affiliations}
	\item 5. Physikalisches Institut, Universit\"{a}t Stuttgart, Pfaffenwaldring 57, 70569 Stuttgart, Germany
	\item Institut f\"ur Theoretische Physik III, Universit\"{a}t Stuttgart, Pfaffenwaldring 57, 70569 Stuttgart, Germany
\end{affiliations}

\begin{abstract}
The coupling of electrons to matter is at the heart of our understanding of material properties such as electrical 
conductivity. One of the most intriguing effects is that electron-phonon coupling can lead to the formation of a Cooper pair out of two repelling electrons, the basis for BCS superconductivity\cite{BCS57}. 
Here we study the interaction of a single localized electron with a Bose-Einstein condensate (BEC) and show that it can excite phonons and eventually set the whole condensate into a collective oscillation. We find that the coupling is surprisingly strong as compared to ionic impurities due to the more favorable mass ratio. The electron is held in place by a single charged ionic core forming a Rydberg bound state. This Rydberg electron is described by a wavefunction extending to a size comparable to the dimensions of the BEC, namely up to 8 micrometers. In such a state, corresponding to a principal quantum number of n=202, the Rydberg electron is interacting with several tens of thousands of condensed atoms contained within its orbit. We observe surprisingly long lifetimes and finite size effects due to the electron exploring the wings of the BEC. Based on our results we anticipate future experiments on electron wavefunction imaging, investigation of phonon mediated coupling of single electrons, and applications in quantum optics.
\end{abstract}
\begin{figure}
	\centering
	\includegraphics[width=89mm]{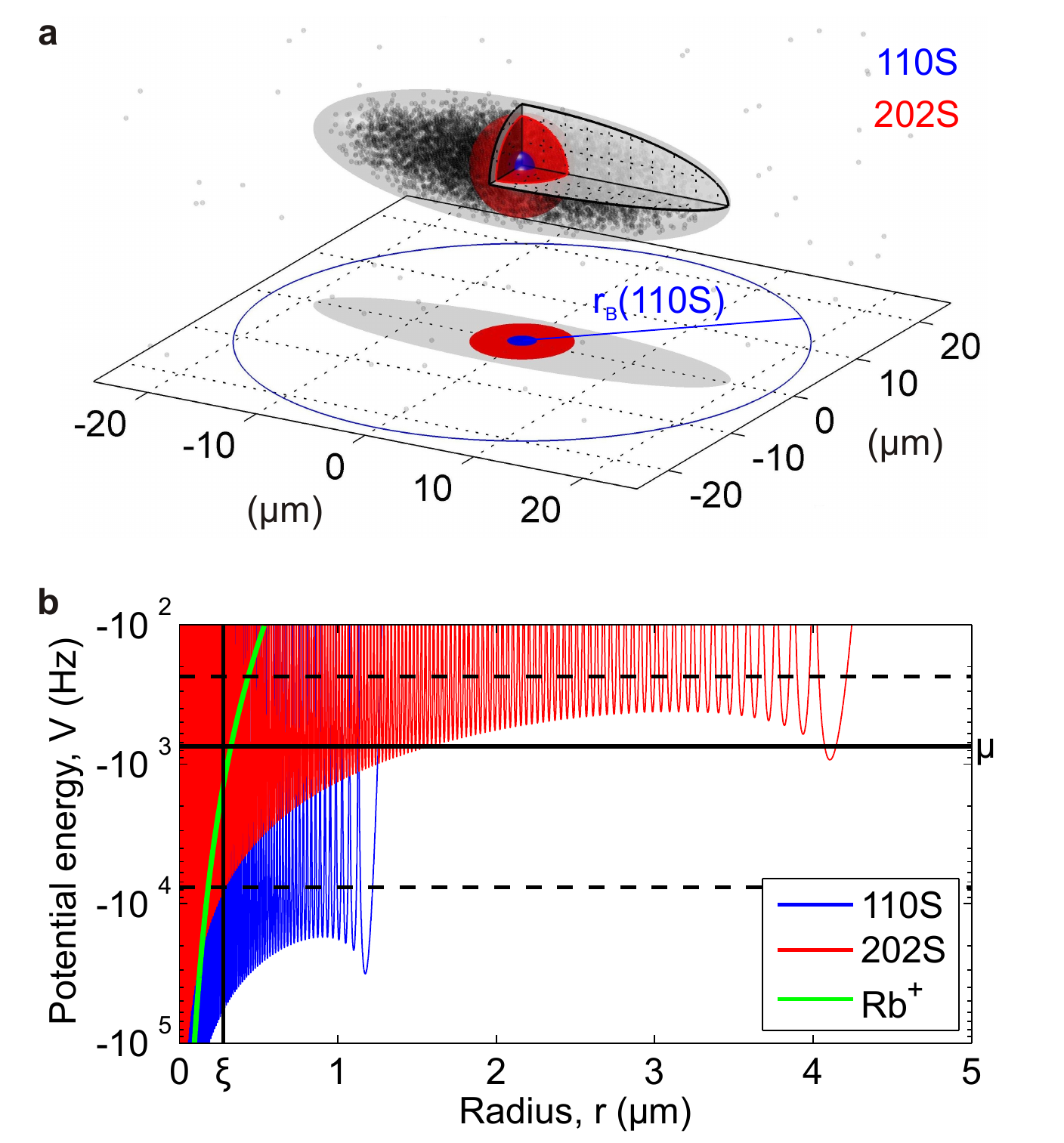} 
	\caption{Size comparison in the spatial and energy domains. \textbf{a} Depending on the Rydberg state the electron impurity is localized in different volumes in a Bose-Einstein condensate consisting of $N=8\cdot10^4$ atoms. The sizes of the lowest (110S, blue) and highest Rydberg state (202S, red) under investigation are indicated. The densities of the BEC and the surrounding thermal cloud are to scale. The lower bound of the blockade radius $r_B$ for the 110S state is denoted as a blue circle in the projection. The blockade radii for the higher Rydberg states are off the scale. \textbf{b} The corresponding interaction potentials $V(r)$  from equation \eqref{eq:molpot} are orders of magnitude stronger than the contribution of the positively charged Rydberg core (green) except for very small distances. The mean interaction strength (black dashed lines) can be set below and above the chemical potential $\text{\textmu}=745\,\text{Hz}$ of the condensate (horizontal black line) by choosing the Rydberg state. The healing length $\xi=274\,\text{nm}$ of the condensate (vertical black line) is much smaller than the spatial extent of the electron wavefunction.
	\label{fig:scheme}}
\end{figure}
\begin{figure}
	\centering
	\includegraphics[width=89mm]{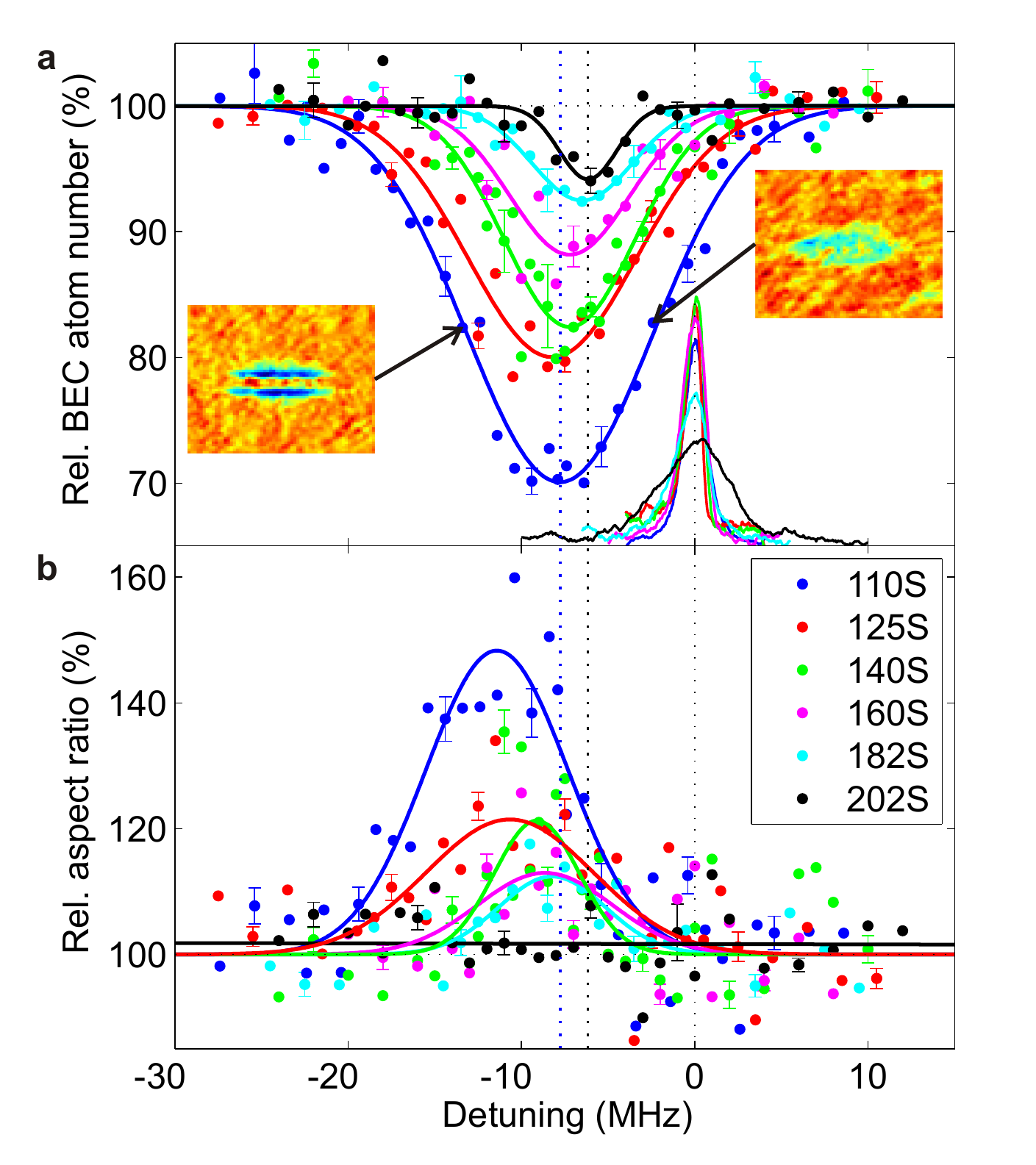}
	\caption{Rydberg excitation spectra for different principal quantum numbers. Relative change of the BEC atom number (\textbf{a}) and aspect ratio (\textbf{b}) after time of flight. The solid lines are Gaussian fits to the data. The zero position is determined as the position of the Rydberg line in the thermal cloud, as shown for reference. The change of aspect ratio is illustrated in the insets of \textbf{a}, which show difference pictures of the condensate at two distinct spectral positions of the 110S state, where the overall losses (blue) are the same. Due to the deformation the condensate gets more elongated, compensating for the atom losses in the center. For the measurements at the two lowest Rydberg states, $n=110$ and $125$ a Rydberg atom was excited 300 times; for the other states 500 repetitions were chosen. The exemplary error bars are the standard deviation from ten independent measurements. 
	\label{fig:lines}}
\end{figure}
\begin{figure}
	\centering
	\includegraphics[width=89mm]{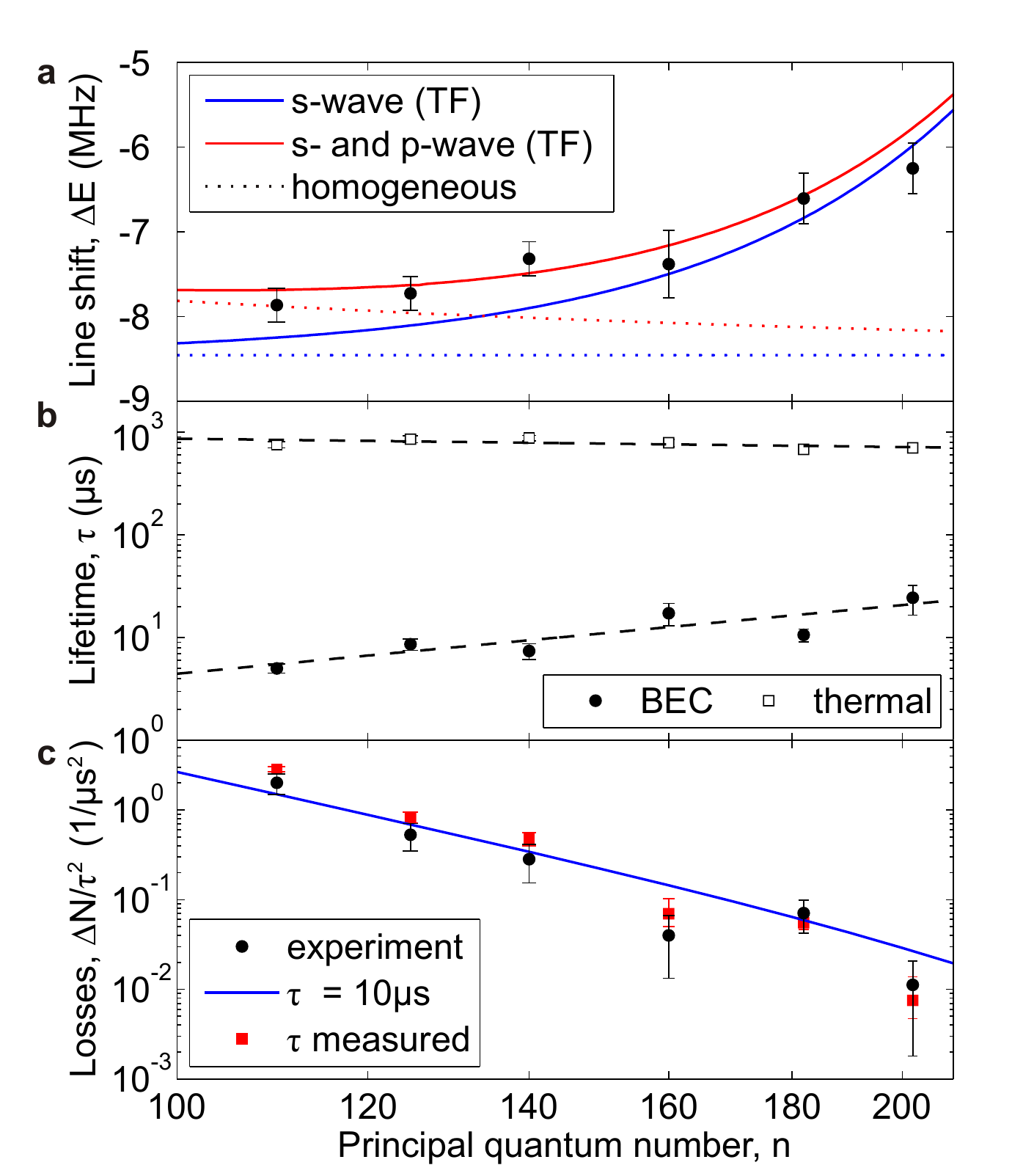}
	\caption{Energy shift and lifetime reduction of the Rydberg state in a condensate and Rydberg electron induced loss of BEC atoms. \textbf{a} Theory curves with constant s-wave scattering length (blue) and taking higher order corrections into account (red) are shown. The corresponding values neglecting the Thomas-Fermi density distribution are indicated as dotted lines. The dashed lines in \textbf{b} are power law fits to measurements in the condensate (circles) and the thermal cloud (squares). In \textbf{c} the atom loss from Fig.~\ref{fig:lines}\textbf{a} is divided by the square of the Rydberg lifetime $\tau$ in the condensate. Theory values taking the measured lifetime into account (red squares) and assuming a constant lifetime of $10\,\text{\textmu s}$ (blue line) are shown. Error bars are $68\%$ confidence bounds of Gaussian fits from Fig.~\ref{fig:lines}\textbf{a} (\textbf{a}), exponential fits to the measured decay (\textbf{b}) and the combination of both (\textbf{c}).
	\label{fig:shifttauloss}}
\end{figure}

Charged impurities were very successfully used as probes for elementary excitations in the early studies of superfluidity in liquid Helium. These applications span the interaction of ion impurities with phonons and rotons\cite{RM60}, the creation and study of vortex lattices by impurities\cite{RR63}, as well as the coupling of electrons to surface ripplons\cite{FHP79}. Additionally they have been proposed for applications in quantum information\cite{PD99}. The emergence of Bose-Einstein condensation of alkali atoms has renewed the interest in impurity physics. Positively charged impurities in a BEC were first created by Penning ionization of metastable atoms\cite{RSB01} and photoionization\cite{CAM02}. Recently single ion trapping\cite{ZPS10} allowed for the first time the study of the interaction between a single charged impurity and a BEC in a well controlled manner. Besides bare losses by  classical scattering, chemical reactions both with the impurity\cite{RZS12} as well as catalytic reactions\cite{HKB12} mediated by the impurity have been observed. Alternatively, the creation of neutral spin impurities in Fermi liquids offers tunable interaction strength with the bulk by means of a Feshbach resonance. A Feshbach resonance allows to change the character of the interaction from free impurities via a quasiparticle state, so called Fermi polarons, to bound diatomic molecules\cite{SSS07,SWS09}. \\ 
For all such impurities, the interaction is inversely proportional to the reduced mass of the impurity and the bulk species\cite{MPS05}. Light impurities such as electrons are therefore, in general, more favorable to obtain strong coupling. However, electrons require an appropriate trapping potential because even the tiniest electric field will lead to a considerable acceleration. The simplest trap for an electron offered by nature is a positively charged nucleus. We use highly excited Rydberg states, where the electron is delocalized over regions of up to several micrometers from the atomic core. At these distances the binding to the core is weak and the Rydberg electron becomes quasi free, making it susceptible to interaction with its environment. 
Nonetheless, the binding to the ionic core is strong enough to provide sufficient trapping of the electron even in a strongly interacting environment. In fact, Amaldi and Segr\`e observed Rydberg absorption series in the range of principle quantum numbers around 30 at pressures reaching one atmosphere\cite{AS34}. In this regime the spatial extend of the electron wavefunction is much larger than the mean interparticle distance, causing large interaction induced line shifts and broadenings. The explanation of these effects led to the nowadays well known Fermi pseudopotential\cite{F34}, describing the short range interaction of the quasi free Rydberg electron at position $\vec{r}$ with neutral ground state atoms at $\vec{R}$:
\begin{equation} \label{eq:pseudopot}
	V_{\text{pseudo}}(\vec{r},\vec{R})=\frac{2\pi\hbar^2a}{m_e}\ \delta(\vec{r}-\vec{R})
\end{equation}
Here the interaction strength is fully characterized by the scattering length $a$. Given the Rydberg electron wavefunction $\Psi(\vec{r})$, the pseudopotential leads to a mean field potential:
\begin{equation} \label{eq:molpot}
	V(\vec{R})=\frac{2\pi\hbar^2a}{m_e}\ \left|\Psi(\vec{R})\right|^2
\end{equation} 
The interaction is restricted to a range given by the size of the Rydberg atom $\propto n^2a_0$, where $a_0$ denotes the Bohr radius. If this is comparable to the mean interparticle distance this interaction can lead to a bound state\cite{GDS00} in the case of negative scattering lengths $a$. Diatomic\cite{BBN09} and triatomic molecules\cite{BBN10} of this class have been experimentally observed for $^{87}$Rb atoms at principal quantum numbers in the range of $n=40$ and densities of the order of $10^{12}\,\text{cm}^{-3}$. At higher densities however, the lifetime of these bound states was found to decrease significantly\cite{BBN11}. \\
The experiments described here are performed in a completely different regime. We combine much higher Rydberg states at principal quantum numbers $n=110-202$ with high densities of up to $10^{14}\,\text{cm}^{-3}$ in a BEC. In this regime, a significant fraction of the  BEC is inside the electron orbit. Consequently, the single electron impurity has an impact on the BEC wavefunction as a whole. We vary the absolute number of atoms inside the electron orbit from 700 to up to 30,000 by changing the principal quantum number which increases the spatial extent of the electron wavefunction up to the wings of the condensate (see Fig.~\ref{fig:scheme}). \\
We start with a magnetically trapped Bose-Einstein condensate of $^{87}$Rb atoms in the $\left|5S_{1/2},\,m_F=2\right>$ state. 
We illuminate the BEC with light from two lasers coupling the ground state to the $\left|nS_{1/2},\, m_S=1/2\right>$ state ($n=110-202$) via the intermediate $5P_{3/2}$ state (see Methods). The strong van-der-Waals interaction between Rydberg atoms in such high states prevents the simultaneous excitation of two Rydberg atoms within a certain distance $r_B$. For Rydberg states with $n>100$ this blockade effect\cite{SWM10} allows only a single Rydberg excitation in the whole BEC, since the blockade radius here is significantly larger than the Thomas-Fermi radii of the condensate (see Fig.~\ref{fig:scheme}\textbf{a} and Supplementary Information). This enables us to create one single electron impurity in the condensate. 
After a defined interaction time on the order of the Rydberg state lifetime, we apply an electric field pulse to remove the ionic core as well as the electron. We repeat this sequence several hundred times in a single condensate, before releasing the BEC and taking an absorption image after $50\,\text{ms}$ time of flight. The long expansion time isolates the low momentum components of the trapped atom cloud, as the free particles have already traveled out of the region of interest.
From these time of flight images we extract the change in atom number and the aspect ratio of the BEC at different detunings of the excitation lasers relative to the single atom Rydberg level and different principal quantum numbers (see Fig.~\ref{fig:lines}). \\   
We observe the impact of the electron impurity as a loss of atoms from the BEC after time of flight. Compared to the Rydberg resonances measured in a thermal cloud at low densities the observed lines in the BEC are shifted to the red inversely proportional to the Rydberg quantum number. We can explain this shift by the low energy scattering of the electron from the BEC atoms. While the mean depth of the interaction potential decreases from $12\,\text{kHz}$ to $290\,\text{Hz}$ with increasing principal quantum number, the number of atoms inside the electron wavefunction increases. Integrating the mean field potential $V(\vec{R})$ over all atoms inside the Rydberg electron wavefunction we obtain the total shift, which depends to lowest order\cite{AS34,F34} only on the mean atomic density $\overline{n}$:
\begin{equation} \label{eq:shift}
	\Delta E=\int_{\text{Ryd}}\!\mathrm{d}^3R\ V(\vec{R})n(\vec{R})=\frac{2\pi\hbar^2a}{m_e}\overline{n}
\end{equation}
For homogeneous densities the shift is therefore independent of the actual Rydberg state. This is consistent with our measurements at smaller principal quantum numbers where the local density approximation is fulfilled. For higher principal quantum numbers however, we find a significant deviation. When the spatial extent of the Rydberg atom becomes comparable to the radial size of the BEC (see Fig.~\ref{fig:scheme}\textbf{a}), the electron is exploring mainly the low density wings of the condensate. We can account for this by averaging the Thomas-Fermi density distribution over the size of the Rydberg atom. We assume the Rydberg atom is being excited in the center of the condensate and use the average value of the peak density $\overline{n}=8.6\cdot10^{13}\,\text{cm}^{-3}$ over one sequence. In Fig.~\ref{fig:shifttauloss}\textbf{a} the calculated shift is plotted in comparison to the line shifts extracted from Fig.~\ref{fig:lines}\textbf{a}. In addition to the values for a constant scattering length\cite{BTF01} of $a=-16.1\,a_0$, a calculation taking higher order scattering theory into account is shown in Fig.~\ref{fig:shifttauloss}\textbf{c} (see Supplementary Information). This simple model without any fit parameters quantitatively agrees remarkably well with our data. \\
The density dependent shift of the Rydberg line allows us to control the position of the electron impurity inside the BEC much better than our optical resolution simply by changing the frequency of the excitation lasers.
This is confirmed by the measured deformation of the condensate (Fig.~\ref{fig:lines}\textbf{b}). The interaction of the electron impurity with the BEC excites a quadrupole shape oscillation that leads to a change in the aspect ratio after time of flight.
In particular, we observe the strongest deformation on the red side of the BEC loss feature, i.e. in the region of largest density-induced shift. 
Here the electron impurity is localized at the center of the condensate and strong deformation of the condensate can be expected, whereas on the blue side and at high principal quantum numbers the mechanical effect averages to zero for repeated excitations. A quantitative modeling of this effect however, is more involved and is subject to future studies. \\
In order to determine the actual interaction time of an individual impurity with the BEC we measure the lifetime of the Rydberg atoms (see Fig.~\ref{fig:shifttauloss}\textbf{b} and Supplementary Information).
From empirical scaling laws\cite{BRT09} one would expect the Rydberg lifetimes $\tau$ to increase with principal quantum number as $n^3$, here from $1.7\,\text{ms}$ ($n=110$) to $10.8\,\text{ms}$ ($n=202$). However, we already observe a reduced lifetime of around $780\,\text{\textmu s}$ at densities around $10^{12}\,\text{cm}^{-3}$ in the thermal cloud for all Rydberg states investigated between $n=110$ and $202$. In the condensate, at peak densities just below $10^{14}\,\text{cm}^{-3}$, we find the lifetime to be further reduced by about two orders of magnitude. In this high density regime we observe a lifetime increasing with principal quantum number. These two observations suggest that there is a dominant decay mechanism which is mainly dependent upon the density of the background gas. Further discussion of this effect can be found in the Supplementary Information. \\ 
We now turn to the quantitative analysis of the fraction of atoms removed from the condensate due to the electron impurity. To fully model this process we have to take the coherent properties of the condensate and its excitations into account.  We use perturbation theory to calculate the number of excitations in the condensate. Expressing the interaction potential $V(\vec{r})$ from equation \eqref{eq:molpot} in terms of Fourier components $\rhoq = \int\!\mathrm{d}^3 r\, e^{-i\vec{q}\vec{r}} |\Psi(\vec{r})|^2$ of the electron density, the relevant part of the interaction reads (see Supplementary Information):
\begin{equation}
	\hat{H}_{\text{int}}=\frac{2\pi\hbar^2 a}{m_e}\,\frac{\bar{n}}{\sqrt{N}}\sum_{\vec{q}\ne 0}  \rhoq(\bogu_{\vecq}-\bogv_{\vecq})(\bopd_{\vecq}+\bop_{-\vec{q}})
\end{equation}
where $\bogu_{\vec{q}}$ and $\bogv_{\vec{q}}$ denote the Bogoliubov factors and $\bopd_{\vec{q}}$ is the creation operator of a collective excitation with quasi momentum $\vec{q}$. The characteristic length scale of the interaction potential is given by the outer edge located at about the Bohr orbit $\propto n^2a_0$. This length scale is larger than the healing length of the BEC for all Rydberg states investigated here (see Fig.~\ref{fig:scheme}\textbf{b}). Therefore, a significant part of the excitations are created in the phonon regime. After time of flight both phonon and free particle excitations can be detected as atom losses. Furthermore, the interaction time of each Rydberg electron with the condensate is finite, limited by its lifetime and the experimental sequence. Taking these corrections into account (see Supplementary Information), we are able to reproduce our experimental results. 
Fig.~\ref{fig:shifttauloss}\textbf{c} shows the maximum atom loss extracted from the data in Fig.~\ref{fig:lines}\textbf{a}, divided by the square of the lifetimes $\tau$ measured for each Rydberg state. This way we remove the main dependency on $\tau$ (see Supplementary Information). The solid line shows the atom loss predicted by our Bogoliubov calculation assuming a constant lifetime of $\tau=10\,\text{\textmu s}$ for all Rydberg states. This already reproduces the overall effect very well. Even better agreement between our data and theory can be achieved if we use the explicitly measured lifetime of each Rydberg state (red points in Fig.~\ref{fig:shifttauloss}\textbf{c}).\\
In view of our results on the coupling of a single electron to BEC excitations further phenomena like trapping of a whole condensate by an impurity come within reach. A repulsive interaction between electron and BEC could be achieved by changing the spin state of the electron. Tiny electric fields are sufficient to deform and manipulate the Rydberg electron wavefunction offering further options to control the coupling. The interaction of the impurity with excitations already present in the condensate could provide a model system for phonon mediated coupling of electrons. From the perspective of the single Rydberg atom the strong interaction of the excited electron with the background gas opens the possibility of manipulating its quantum mechanical state as well as its motional degrees of freedom. These effects enable intriguing quantum optics applications. For example, the scattering from the background gas could serve as a source of dephasing, forming a crucial part in the proposal for a single photon absorber\cite{HLW11}. Finally, the BEC provides a sensitive probe for individual Rydberg atoms since the depletion of the BEC is localized inside the volume of the electron wavefunction. Highly ordered quasi crystalline samples of multiple Rydberg atoms in one condensate or even two overlapping Rydberg atoms could be created and probed by means of double resonance spectroscopy\cite{RYL08}. Making use of well established techniques like in-situ phase contrast imaging, even the imaging of a single electron orbital seems feasible.     

\begin{methods}
We start with a condensate of $N=8\cdot10^4$ $^{87}$Rb atoms in the $\left|5S_{1/2},\,m_F=2\right>$ state in a cloverleaf type magnetic trap at a high magnetic offset field of $1.355\,\text{mT}$ (radial/axial trap frequencies $\omega_{\rho}=2\pi\cdot81.7\,\text{Hz}$ and $\omega_{z}=2\pi\cdot22.4\,\text{Hz}$). We excite Rydberg states $\left|nS_{1/2},\,m_S=1/2\right>$  ($n=110-202$) via a two-photon transition detuned by $500\,\text{MHz}$ from the intermediate $5P_{3/2}$ state using continuous wave diode lasers at $780\,\text{nm}$ and $480\,\text{nm}$. The blue laser beam has a power of $100\,\text{mW}$ and is focused down to a size of $60\,\text{\textmu m}$ ($1/e^2$ diameter). We choose the power of the red beam typically in the range of $3\,\text{\textmu W}$ at a diameter of $500\,\text{\textmu m}$ ($1/e^2$). The two laser beams are shone in counterpropagating along the magnetic field axis of the trap. We address the desired transition by choosing $\sigma^+$ and $\sigma^-$ polarization for the $780\,\text{nm}$ and  $480\,\text{nm}$ laser respectively. \\
In each condensate a sequence consisting of a $1\,\text{\textmu s}$ light pulse for Rydberg excitation and a $2\,\text{\textmu s}$ electric field pulse for removal of any remaining Rydberg atoms or ions, separated by $10\,\text{\textmu s}$ delay time, is repeated 300-500 times at a rate of $62.5\,\text{kHz}$. The clearance field is set to $5.7\,\text{V/cm}$, well above the ionization threshold of all Rydberg states under investigation. After a time of flight of $50\,\text{ms}$ we take an absorption image of the condensate. We determine the relative change in atom number and aspect ratio by comparing each measurement with a consecutive reference image where the blue Rydberg laser is detuned by more than $40\,\text{MHz}$. \\
Further details about the setup and the data analysis can be found in ref. \citenum{LWN12} and in the Supplementary Information.   
\end{methods}

\bibliographystyle{naturemag}
\bibliography{script}

\begin{addendum}
 \item[Acknowledgements] We would like to thank K. Rz\k{a}\.{z}ewski and J. Hecker Denschlag for valuable discussions and C. Tresp for setting up the Rydberg laser system. This work is funded by the Deutsche Forschungsgemeinschaft (DFG) within the SFB/TRR21 and the project PF~381/4-2. We also acknowledge support by the ERC under contract number 267100 and A.G. acknowledges support from E.U. Marie Curie program ITN-Coherence 265031.
\end{addendum}


\onecolumn

\renewcommand{\figurename}{Supplementary Figure}
\setcounter{figure}{0}

{\huge Supplementary Information} \\  

\section{Thermal cloud data} 
The reference measurements of the unperturbed Rydberg state in the thermal cloud were performed with a sample of $2\cdot10^6$ atoms at a temperature of $2.6\,\text{\textmu K}$. The spectra were taken by laser excitation and subsequent field ionization and ion detection\cite{LWN12}. The electric field for ionization was $5.7\,\text{V/cm}$ throughout all measurements, which is well above the classical ionization threshold of $2.5\,\text{V/cm}$. A complete spectrum was taken in one atomic sample by repeating the cycle of excitation and detection 401 times at a rate of $167\,\text{Hz}$ while varying the laser detuning on each shot. 
The lines shown in Fig.~2\textbf{a} were obtained by averaging 20 spectra. For these spectra a long excitation pulse length of $100\,\text{\textmu s}$ and low red powers of few nW were chosen to minimize linewidth. This linewidth, increasing from just below $1\,\text{MHz}$ at $n=110$ to about $5\,\text{MHz}$ at $n=202$, is clearly limited by the electric field control. We compensated stray electric fields by repeatedly taking Stark parabola along three axes. We estimate the level of electric field control to be on the order of $1\,\text{mV/cm}$. The measured linewidth increased at slightly higher temperatures of the atomic sample, indicating the presence of residual electric field gradients. We expect much smaller line broadening caused by inhomogeneous fields in the BEC due to its smaller spatial extend. \\
For the lifetime measurements in Fig.~3\textbf{b} spectra with varying time delay between excitation and detection were taken. Here the same power as for the corresponding condensate measurements was used with pulse lengths of 1 to $4\,\text{\textmu s}$ to obtain a constant signal amplitude for all Rydberg states. For each delay time 5 spectra were averaged and the amplitude extracted by a Gaussian fit.
  
\section{BEC data}
To study the effect of a single Rydberg electron on the condensate a sequence consisting of a $1\,\text{\textmu s}$ long Rydberg excitation pulse and a $2\,\text{\textmu s}$ long field ionization pulse, with a fixed delay time of $10\,\text{\textmu s}$ in between, was repeated 300-500 times at a rate of $62.5\,\text{kHz}$. The red power was adjusted for each principal quantum number in the range of around $3\,\text{\textmu W}$ to the plateau value of a saturation curve\cite{HRB07}. After a time of flight of $50\,\text{ms}$ an absorption image was taken from which the atom number and aspect ratio of the BEC was extracted. We determine the atom number by summing all pixels in a rectangle around the condensate and normalizing the number for each picture onto the average background signal in a region without atoms. The aspect ratio was obtained from a one dimensional Thomas-Fermi fit to slices integrated along 11 pixels (pixel size $6.45\,\text{\textmu m}$) along the long and short axis of the condensate. \\
During each sequence, the number of atoms in the condensate decreased from $8\cdot10^4$ to around $5\cdot10^4$ atoms, even in the absence of Rydberg excitations, mainly due to off-resonant scattering from the intermediate $5P_{3/2}$ state and heating. To eliminate this effect from the data, as well as to reduce the effect of drifts in atom number and deformation during time of flight originating from residual magnetic field gradients, the data from each absorption image was related to a reference measurement taken immediately before or after where the blue Rydberg laser was detuned by more than $40\,\text{MHz}$. Therefore any Rydberg excitation was avoided while keeping the loss due to the red Rydberg laser and the deformation due to the focused blue Rydberg laser constant. For each data point in Fig.~2 ten measurements were averaged. For the insets in Fig.~2 we subtracted the reference image from the image with Rydberg excitation and averaged again over ten repetitions. Note that taking one BEC spectrum as in Fig.~2 takes therefore at least ten hours of uninterrupted measurement time without preparation and warming up, while an averaged spectrum in the thermal cloud can be obtained easily within ten minutes at much better resolution. We monitored the initial BEC atom number during the experiments and it turns out that small drifts are negligible except for the measured line shift of the Rydberg state. Therefore we normalize the line positions extracted from Fig.~2\textbf{a} to the mean peak density $\propto N^{2/5}$ to obtain the values shown in Fig.~3\textbf{a}.   
\\
For the lifetime measurements in Fig.~3\textbf{b} the delay time between excitation and ionization was varied. The detuning of the Rydberg lasers was set to the position of maximum atom loss in Fig.~2. The overall length of the sequence was adapted to the maximum delay time but kept constant throughout the measurement of one Rydberg state. For the longest delay times we found no change of the measured BEC atom losses if we switched the field ionization off. From this we can conclude that we really detect the decay of the Rydberg atom itself and not only the decay of its effect on the condensate. 

\section{Effect of impurity atom mass ratio on interaction strength}
In the main paper we state that light impurities in general cause a stronger interaction with the bulk. A thermodynamic consideration based on ref. \citenum{MPS05} provides an estimate of the excess number of atoms $\Delta N$ which is accumulated around a single impurity in equilibrium. This number can be expressed in terms of the reduced masses $m$ and scattering lengths $a$ for the atom-atom ($aa$) and atom-impurity ($ai$) scattering respectively:   
\begin{equation}
	\Delta N=-\frac{m_{aa}}{m_{ai}}\frac{a_{ai}}{a_{aa}}
\end{equation}
The atom-impurity scattering length $a_{ai}$ itself is also a function of the reduced mass $m_{ai}$. The order of magnitude for a singly charged impurity interacting with atoms with polarizability $\alpha$ can be estimated as the characteristic radius $r_{ai}$ of the polarization potential\cite{ZPS10}:
\begin{equation}
	r_{ai}=\sqrt{\frac{m_{ai}\alpha e^2}{(4\pi\epsilon_0\hbar)^2}}
\end{equation}
The absolute value of $r_{ai}=18.0\,a_0$ agrees quite well with the actual value of $a_{ai}=-16.1\,a_0$ for the $e^-$-$^{87}$Rb triplet scattering observed in this paper, whereas the singlet scattering length $a_{ai}=0.627\,a_0$ differs considerably\cite{BTF01}. This leads to a scaling of $\Delta N\propto1/\sqrt{m_{ai}}$. In case of $^{87}$Rb one can therefore expect the effect of an electron impurity to be about a factor of $400$ stronger than that of a positively charged ion of the same element as the bulk atoms. The value of $\Delta N$ then exceeds one thousand. Here we are clearly reaching the limits of weak perturbation assumed in the derivation of $\Delta N$.   

\section{Blockade radius} 
The interaction between two Rydberg atoms in an $S$ state is a purely repulsive van-der-Waals interaction $C_6/r^6$ except for very small interatomic distances $r$. In the main paper we state the minimum distance between two Rydberg atoms which can still be excited with one laser at fixed center frequency $f$ but with finite linewidth $\Delta f$, the blockade radius $r_B$. As a simple estimate for the blockade radius we equate the van-der-Waals interaction with the linewidth $\Delta f$ and obtain:
\begin{equation}
	r_B\approx\sqrt[6]{C_6/(h\Delta f)}
\end{equation}
As discussed above, this linewidth is dominated by electric field gradients which are expected to play a much smaller role on BEC length scales. Therefore, the actual blockade radius around one Rydberg atom in the condensate is expected to be much larger than estimated. However, we note that this simple picture of the Rydberg blockade in our case has to be extended by the density dependent line shift according to equation (3) which can cause an antiblockade effect. Since the van-der-Waals interaction is purely repulsive a Rydberg excitation in the low density region of the condensate, at a detuning close to zero with respect to the thermal cloud, could possibly tune a second excitation more in the center of the BEC into resonance. The zero crossings of the combined interaction and density dependent potentials, therefore, in principle allow to create quasi crystalline ordered structures of Rydberg atoms. More important for the work presented here, however, is that for large detunings Rydberg atoms are excited preferentially in the center of the condensate. In this case the density gradient and the Rydberg-Rydberg interaction are of the same sign and there is consequently no antiblockade effect present. On the contrary, the blockade by a single excitation here is even more effective. 

\section{Rydberg decay} 
We observe shorter lifetimes of Rydberg states than expected from spontaneous decay only (see Fig.~3\textbf{b}). Furthermore, the comparison between measurements in the thermal cloud and in the condensate suggest a linear dependence on the density of ground state atoms. Initially one could therefore expect that the weak binding of the highly excited Rydberg electron, orbiting micrometers away from the nucleus, could be destroyed immediately in a scattering event from a heavy object like a ground state atom. However, energy conservation would require a transition to a neighboring Rydberg state. As the closest states are at least a few GHz apart, orders of magnitude more than the maximum classical energy transfer in such a scattering event, these processes are very unlikely to happen\cite{H79}. The quantization of the Rydberg states here leads to a stabilization of the Rydberg electron at high density. We also verified experimentally that there is no stepwise decay of the Rydberg electron back to the ground state. To this end we measured the BEC atom loss with a $2\,\text{\textmu s}$ long electric field pulse of variable strength directly after Rydberg excitation and after a $3\,\text{\textmu s}$ wait. At the end of each sequence we apply another electric field pulse which removes any possibly remaining Rydberg atoms. The sequence is illustrated in the inset of Supplementary Fig.~S\ref{fig:checkfordecay}. 
\begin{figure}
	\centering
	\includegraphics[width=89mm]{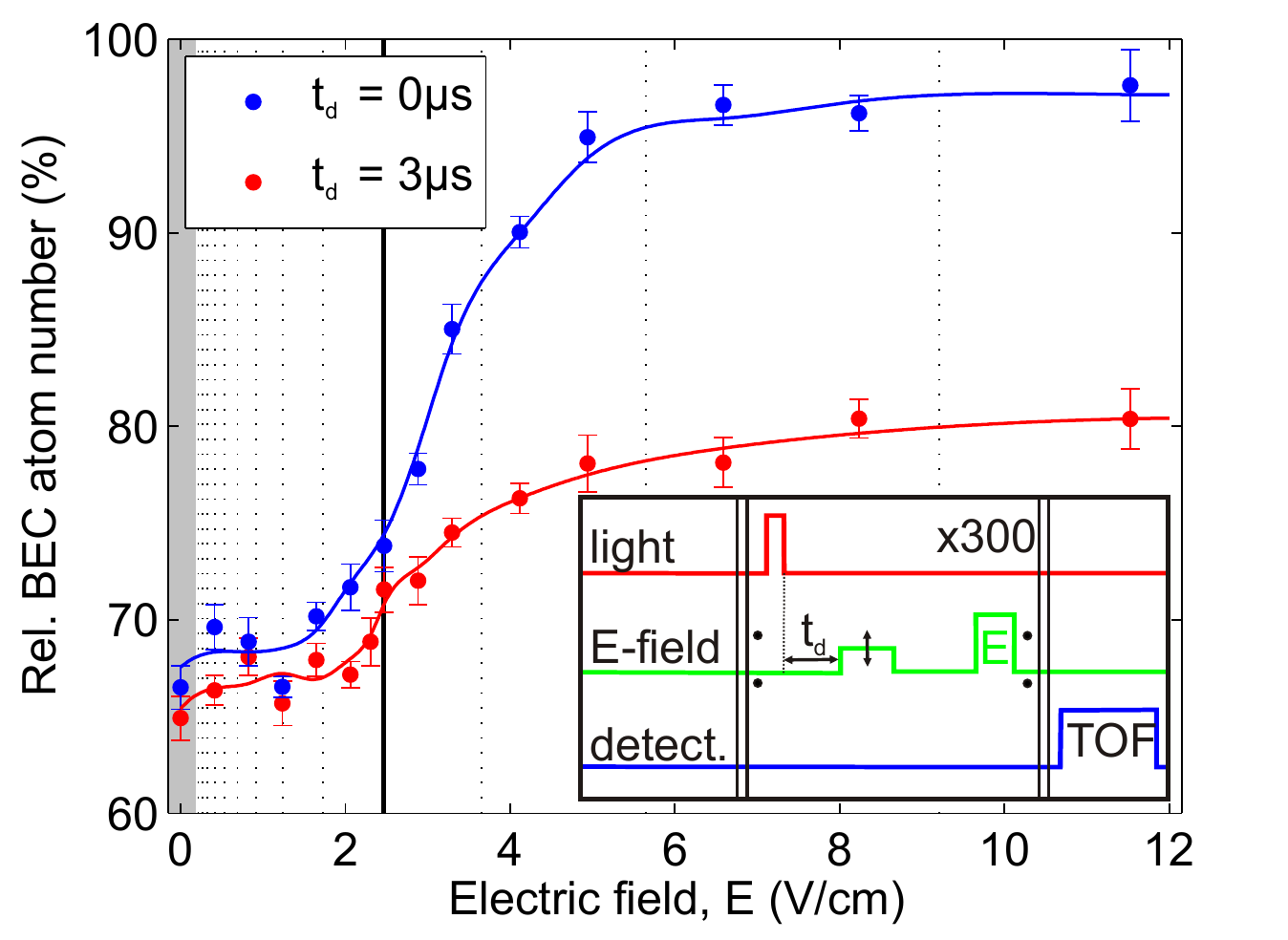} 
	\caption{Study of Rydberg decay of the 110S state. The BEC atom loss was measured varying the electric field strength of pulses with a delay time of $t_d=3\,\text{\textmu s}$ (red) and immediately after Rydberg excitation (blue). The dotted vertical lines indicate the classical ionization threshold for Rydberg states in steps of ten principal quantum numbers. The solid black line corresponds to the value for $n=110$. In the grey shaded area the electric field is not strong enough to extract a possibly existing ion from the condensate during the electric field pulse. The experimental sequence is depicted in the inset.
	\label{fig:checkfordecay}}
\end{figure}
We found that the BEC atom loss starts to vanish at a distinct electric field strength, which is on the order of the classical field ionizing threshold and does not depend on the delay time. From this we can infer that during this $3\,\text{\textmu s}$, which is a considerable fraction of the measured lifetime of $(5.0\pm0.5)\,\text{\textmu s}$, the Rydberg atom is not decaying to states with significantly lower principal quantum numbers. This means that the Rydberg atoms are either decaying directly to relatively low states (as they do for spontaneous decay) or into the continuum. Possible processes could be associative ionization (so called Hornbeck-Molnar ionization) or ion pair formation. In the first process, the Rydberg atom and a neutral ground state atom form a positively charged $Rb_2^+$ molecular ion; in the latter, a pair of positive and negative Rubidium ions are produced\cite{BDC86}. 

\section{Higher order terms in scattering potential} 
For simplicity the discussion in the main paper is restricted to a constant scattering length $a$ for the interaction of the slow Rydberg electron with the ground state atom. However, as indicated in Fig.~3\textbf{a} there is some correction depending on the momentum $k(R)$ of the electron at distance $R$ from the Rydberg core. The contribution to the s-wave scattering length $a$ is proportional to the polarizability $\alpha$ of the ground state atoms:
\begin{equation}
	a(k(\vec{R}))=a+ \frac{\hbar^2}{e^2a_0^2m_e}\cdot\frac{\pi}{3} \alpha k(\vec{R})+ O(k^2)
\end{equation}
To the same order of approximation, the p-wave scattering potential is\cite{O77}:
\begin{equation}
	V_p(\vec{R})=-\frac{e^2}{4\pi\epsilon_0}\cdot\frac{2\pi^2}{5}\frac{\alpha}{k(\vec{R})}\left|\vec{\nabla}\Psi(\vec{R})\right|^2
\end{equation}
We obtained the theoretical prediction shown in Fig.~3\textbf{a} by numerically integrating over the s- and p-wave potentials using a semiclassical approximation for the electron momentum\cite{BBN09}: 
\begin{equation}
	k(R)=\sqrt{\frac{2m_e}{\hbar^2}\left(-\frac{Ryd}{(n-\delta_0)^2}+\frac{1}{(4\pi\epsilon_0)}\frac{e^2}{R}\right)}
\end{equation}
with the Rydberg constant $Ryd$, the quantum defect $\delta_0$, and the vacuum permittivity $\epsilon_0$. As the principal quantum number $n$ of the Rydberg state increases the momentum independent approximation gets better since the mean momentum of the electron is decreasing with $1/n$. Nevertheless, the correction is slightly larger than the error bars of our measurements.
However, discrepancies on this order could be equally explained by systematic errors of the measured peak density or the inhomogeneous line broadening due to Rydberg excitation not exactly at the center of the BEC.      
\section{Coupling to BEC excitations}
\begin{figure}[h]
	\centering
	\includegraphics[width=89mm]{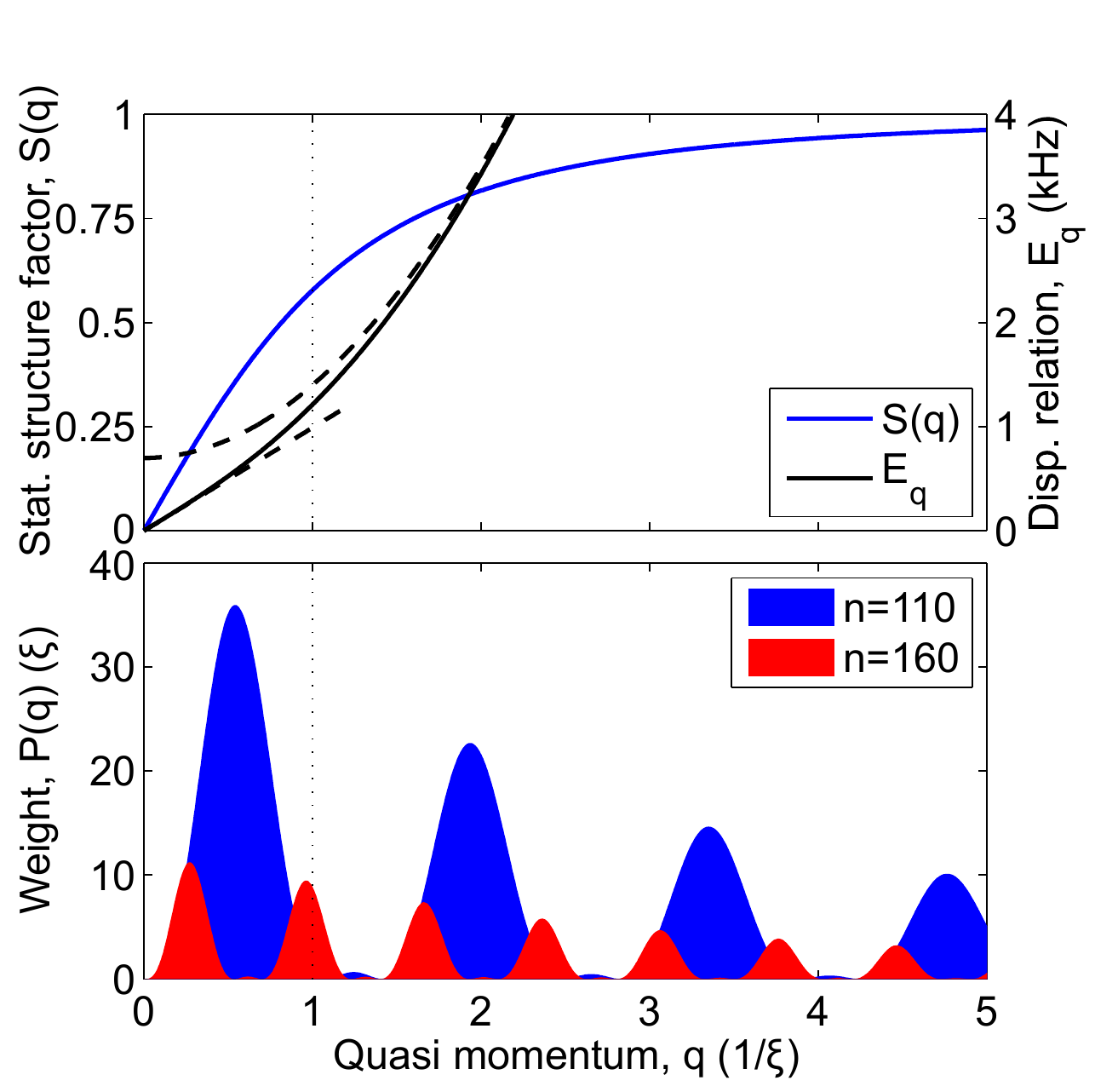} 
	\caption{Nature of excitations in the condensate. In the upper panel, the static structure factor $S(q)$ of the BEC (blue) is shown together with the Bogoliubov dispersion relation $E_q$ (black), which is shown as a reference. The linear and quadratic regimes are indicated as dashed lines. The lower panel shows the total weight $P(q)$ of excitations at different momenta $q$ for two principal quantum numbers $n=110$ and $160$.}
	\label{fig:modes}
\end{figure}

In the s-wave approximation, the interaction between the electronic density $\rho(\vecr)=\absvsq{\Psi(\vecr)}$ in the Rydberg $n$S state and the ground state atoms is described by the interaction potential $V(\vecr)=g \rho(\vecr)$, where $g=2\pi \hbar^2 a/m_e$. As the extent of the wavefunction is smaller than the condensate, we can treat the BEC in the thermodynamic limit, assuming a constant atomic density $n(\vecr)=V^{-1} \sum_{\vecp,\vecq} \ef{i\vecq\vecr}\aopd_{\vecp+\vecq} \aop_{\vecp}$, where $\aop_{\vecp}$ is an annihilation operator of a mode $\vecp$ within the quantization volume $V$. The interaction can therefore be expressed as a convolution in momentum space
\begin{align}
\hat{H}_{\text{int}} = g\intvol n(\vecr)\rho(\vecr) =\frac{g}{V} \sum_{\vecp,\vecq} \aopd_{\vecp+\vecq} \aop_{\vecp} \rhoq
\end{align}
Using the mean field approximation $\aop_0\approx\aopd_0\approx\sqrt{N_0}$, with $N_0$ denoting the number of atoms in the condensate mode, we can write the interaction in terms of Bogoliubov operators $\bop_{\vecq}=\bogu_q \aop_{-\vecq} - \bogv_q \aopd_{\vecq}$ as
\begin{align}
\hat{H}_{\text{int}}\approx\frac{g\sqrt{N}}{V} \sum_{\vecq\ne 0} \rho^{\phantom\dagger}_{\vecq} \big(\bogu_{q}-\bogv_{q}\big) \big(\bopd_{\vecq}+\bop_{-\vecq}\big)
\end{align}
where we have neglected constant energy shifts and higher order corrections. Note that within this order of approximation we can replace $N_0$ by the total atom number $N$. 
To estimate the number of excitations induced by the presence of the Rydberg electron with lifetime $\tau=1/\gamma$ we first consider the probability to excite a certain mode with quasi momentum~$\vecq$ when a perturbation of the type $\hat{H}_{\text{int}} \ef{-\gamma t}$ is applied. To lowest order we have
\begin{align}
P_{0\rightarrow\vecq}&=\Bigg|-\frac{i}{\hbar}\integralb{0}{\infty}{t}\ef{i\omega_{q}t-\gamma t}\braketop{\vecq}{\hat{H}_{\text{int}}}{0}\Bigg|^2
\end{align}
where the initial state $\ket{0}$ denotes the BEC ground state within Bogoliubov approximation and the final state is given by $\ket{\vecq}=\bopd_{\vecq}\ket{0}$, which is an excited state with energy $E_{q}=\hbar\omega_{q}=\sqrt{\epsilon_{q}^2+2\overline{n}g_c\epsilon_{q}}$. Here we have used the recoil energy $\epsilon_{q}=\hbar^2 q^2/2m_\text{Rb}$, the mean density $\overline{n}=N/V$ and the atom-atom coupling constant $g_c=4\pi \hbar ^2a_{\text{Rb}}/m_{\text{Rb}}$ with the s-wave scattering length $a_{\text{Rb}}$. 
For the probability we now find
\begin{align}
P_{0\rightarrow\vecq}=\frac{g^2 \rho_{\vecq}^2 }{V^2\hbar^2}   \integral{\omega}S(\vecq,\omega)\absvsq{C(\omega)} = \frac{g^2 \rho_{\vecq}^2}{V^2\hbar^2} N \frac{\epsilon_{q}}{E_{q}} \absvsq{C(\omega_q)}
\end{align}
where $S({\vecq},\omega)=N \,\epsilon_{q}/E_{q} \cdot \delta(\omega-\omega_{q})$ is the dynamic structure factor of the BEC and $C(\omega)=1/(\gamma-i \omega)$ is the Fourier transform of the exponential decay.
During the time of flight process, the atom-atom interactions quickly become negligible and the Bogoliubov modes are converted into free particles. Using $N=\sum_{\vecp} \aopd_{\vecp}\aop_{\vecp}$, we find that $\braketop{\vecq}{N}{\vecq}-\braketop{0}{N}{0} = u_{q}^2+v_{q}^2$ additional particles are in the excited state. The total number of lost atoms $\Delta N$ can now be expressed as
\begin{align}
\Delta N &= \sum_{\vecq} P_{0\rightarrow\vecq}\, (u_{q}^2+v_{q}^2)
\end{align}
Replacing the sum by an integral and using the healing length $\xi=1/\sqrt{8\pi \overline{n} a_{\text{Rb}}}$, we have
\begin{align} \label{eq:lostpertau2}
\Delta N/\tau^2&= \frac{1}{2\pi^2}\frac{\overline{n} g^2}{\hbar^2} \integral{q} q^2\rho_q^2\,\frac{1+(q\xi)^2}{2+(q\xi)^2} \, \frac{1}{1+\omega_q^2/\gamma^2}
\end{align}
where we have separated the main dependency on the two measured quantities on the left hand side.
Fig.~S\ref{fig:modes} illustrates the nature of the excitations that are generated by the Rydberg electron. The static structure factor $S(\vec{q})=\epsilon_q/E_q$ suppresses excitations at low momenta $q$. Nonetheless, the excitation weight $P(q)\sim P_{0\rightarrow q} q^2$ shows a clear maximum located at $q\approx 2/R_e < 1/\xi$, where $R_e=2a_0n^2 $ is the radial extent of the Rydberg electron wavefunction. This lies well in the phonon regime for all principal quantum numbers investigated in the experiment.
Due to the $q^2$ factor from the spherical integration there is also a sizable contribution of free particle excitations, which appear as equally spaced peaks with decreasing magnitude.\\ 
Some experimental details require extensions to equation \eqref{eq:lostpertau2}. First, to account for density inhomogeneities due to the external potential in a simple way we integrate the Thomas-Fermi density profile over the volume which is enclosed by the Rydberg electron, a sphere of radius $R_e$ placed in the center of the atomic cloud. The effective density is then given by the mean value $\overline{n}=\bb{1-(2R_e/5R_\rho)^2 - (R_e/5R_z)^2} n_0$ on this sphere, where $n_0$ is the peak density and $R_\rho$ ($R_z$) is the Thomas-Fermi radius of the cigar shaped BEC in radial (axial) direction.
Second, in the experimental sequence, the interaction between the Rydberg electron and the ground state atoms is suddenly terminated after a certain time $t_c$ at which the field ionization occurs. To account for this, the function $C(\omega)$ is modified accordingly:
\begin{align}
\absvsq{C(\omega)}=\Bigg| \integralb{0}{t_c}{t}\ef{i\omega t-\gamma t} \Bigg|^2=\frac{1+\ef{-2\gamma t_c} -2 \ef{-\gamma t_c} \cos(\omega t_c)}{\gamma^2+\omega^2}
\end{align}
The final correction concerns the way the losses are detected in the experiment. In the absorption images, excitations at small momenta are not distinguished from the condensate fraction due to finite momentum components in the Thomas-Fermi profile. A lower cutoff may thus be introduced in the radial $q$ integration. It turns out that this correction is small and almost all excitations will be detected as losses.  
  
\end{document}